\documentclass[twocolumn,prl,preprintnumbers]{revtex4}
\usepackage[applemac]{inputenc}
\usepackage[T1]{fontenc} 
\usepackage{lipsum}
\usepackage{amsmath} 
\usepackage{amsfonts} 
\usepackage{amssymb}
\usepackage{graphicx} 
\usepackage[usenames]{color} 

\usepackage{bbm}
\def\beq{\begin{equation}}
\def\eeq{\end{equation}}
\def\bsp{\begin{split}}
\def\esp{\end{split}}
\def\bea{\begin{eqnarray}}
\def\eea{\end{eqnarray}}
\def\ba{\begin{array}}
\def\ea{\end{array}}

\def\l.{\left.}
\def\r.{\right.}

\def\part{\partial}

\begin{document}
\title{Macroscopic quantum tunneling of two coupled particles in the presence of a transverse magnetic field.}
\author{Solomon Akaraka Owerre}
\affiliation{Groupe de physique des particules, D\'epartement de physique,
Universit\'e de Montr\'eal,
C.P. 6128, succ. centre-ville, Montr\'eal, 
Qu\'ebec, Canada, H3C 3J7 }

\begin{abstract}

\section{Abstract} Two coupled particles of  identical masses but opposite charges, with a constant transverse external magnetic field and an external potential, interacting with a bath of harmonic oscillators are studied. We show that the problem cannot be mapped to a one-dimensional problem like the one in Ref. \cite{pa}, it strictly remains two-dimensional. We calculate the effective action both for the case of linear coupling to the bath and without a linear coupling using imaginary time path integral at finite temperature. At zero temperature we use Leggett's prescription to derive the effective action. In the limit of zero magnetic field we recover a two dimensional version of the result derived in Ref.  \cite{em1} for the case of two identical particles. We find that in the limit of strong dissipation, the effective action reduces to a two dimensional version of the Caldeira-Leggett form in terms of the reduced mass and the magnetic field. The case of Ohmic dissipation with the motion of the two particles damped by the Ohmic frictional constant $\eta$ is studied in detail. 

\end{abstract}


\maketitle

\section{Introduction}
Macroscopic quantum tunneling with dissipation has become the subject of interest in quantum statistical mechanics and condensed matter physics for many years \cite{al,cl,cl1}. This mainly involves the influence of the environment (thermal bath of harmonic oscillators) on the tunneling of a macroscopic particle with variable, say $q$, out of an external potential $V(q)$, which is assumed to have a metastable minimum. In most cases of physical interest, it is assumed that $q$ interacts linearly with the environmental coordinate say $x_{\alpha}$ ($\alpha  = 1,2,\cdots$) at a certain temperature $T$. The breakthrough in this subject was made by Caldeira and Leggett \cite{cl1}. They considered a Euclidean Lagrangian of the form
\bea
 \begin{split}
\mathcal{L}_E&= \frac{1}{2}M \dot{q}^2+ V(q) + \sum_{\alpha}\frac{1}{2}m_{\alpha}(\dot{x}_{\alpha}^{2} + \omega_{\alpha}^2 x_{\alpha}^2)\\&+q\sum_{\alpha}c_{\alpha}x_{\alpha},
 \end{split}
\eea
where the parameters $m_{\alpha}, \omega_{\alpha}, c_{\alpha}$ need not to be known in detail.
The partition function is given by 
\bea
\begin{split}
K(q,x_{\alpha};\tau)& =\int \mathcal{D}q(\tau) \int  \prod_{\alpha}\mathcal{D}x_{\alpha}(\tau)\exp\left(-S_E\right),
\end{split}
\eea
where 
\bea
S_E = \int_{0}^{\tau}d\tau \mathcal{L}_E. 
\eea
Performing the functional integral over $x_{\alpha}$ in the limit $\tau \rightarrow \infty$ gives
\bea
K(q;\tau) =\int\mathcal{D}q(\tau)\exp\left(-S_E^{eff}\right),
\eea
where the effective action is given by
\bea
\begin{split}
S_E^{eff}&= \int_{0}^{\tau}d{\tau}\left[\frac{1}{2}M\dot{q}^2 +V(q)
\right] \\& +\frac{\eta}{4\pi}\int_{-\infty}^{\infty}d\tau^{\prime}\int_{0}^{\tau}d{\tau} \frac{\left[q(\tau)-q(\tau^{\prime})\right]^2}{(\tau-\tau^{\prime})^2},
\end{split}
\label{5}
\eea

and $\eta$ is the frictional constant.
Chudnovsky \cite{em1} generalized this formalism by considering two macroscopic particles that interact with each other via a nonlinear potential $V(\lvert x_1- x_2\rvert)$ with the coordinate $x_2$ linearly coupled to the environment. The Euclidean Lagrangian is of the form
\bea
 \begin{split}
\mathcal{L}_E&=\frac{1}{2}M_1\dot{x}_1^2 +\frac{1}{2}M_2\dot{x}_2^2 +V(\lvert x_1- x_2\rvert)+ \frac{1}{2}\sum_{\alpha}m_{\alpha}\dot{x}_{\alpha}^{2}\\&+ \frac{1}{2}\sum_{\alpha}m_{\alpha}\omega_{\alpha}^2 (x_{\alpha}-x_2)^2 .
\end{split} 
\eea 
Integrating out the environmental degree of freedom and using the new coordinates
\bea
\begin{split}
q &= x_1-x_2,\\
r&= \frac{M_1 x_1+M_2 x_2}{M_1 +M_2},
\label{10}
\end{split}
\eea
he found that in the limit $M_1\rightarrow \infty$, the effective action reduces to the form of Caldeira and Leggett:
\bea
\begin{split}
S_E^{eff}&= \int_{0}^{\hbar/T}d{\tau}\left[\frac{1}{2}M_2\dot{q}^2 +V(q)
\right] \\& +\frac{1}{2}\int_{-\infty}^{\infty}d\tau^{\prime}\int_{0}^{\hbar/T}d{\tau} \alpha(\tau-\tau^{\prime})\left[q(\tau)-q(\tau^{\prime})\right]^2,
\end{split}
\label{5a}
\eea 
where
\bea
\begin{split}
\alpha (\tau)&= \frac{1}{4}
\sum_{\alpha} m_{\alpha}\omega_{\alpha}^3 \exp(-\omega_{\alpha}\lvert \tau\rvert).
\end{split}
\eea
In this paper, we will generalize Chudnovsky's idea by considering two coupled macroscopic particles, in the presence of
a constant transverse external magnetic field and
an external potential. Due the presence of an external magnetic field, this problem is at least two dimensional.

\section{Model}
In the presence of a magnetic field $\bold{B}$ derivable from a vector potential ($\bold{B}= \boldsymbol{\nabla} \times \bold{A}$), the Euclidean Lagrangian we will first consider is of the form
 \bea
\begin{split}
\mathcal{L}_{E}&= \frac{m_1}{2}\lvert \dot{\bold{x}}_1\rvert^2 + \frac{m_2}{2}\lvert \dot{\bold{x}}_2\rvert^2+  ie\left( \dot{\bold{x}}_1\cdot\bold{A}_1 -  \dot{\bold{x}}_2\cdot\bold{A}_2 \right) \\ &+ \frac{1}{2}m_1\omega^2_1\bold{x}_1^2 + \frac{1}{2}m_2\omega^2_2\bold{x}_2^2+ V(\lvert \bold{x}_1-\bold{x}_2\rvert).
\label{6}
\end{split}
\eea 
Here the vectors have two components given by $\bold{x}_1 = x_{1}^{i} = (x, y)$ and $\bold{x}_2 = x_{2}^{i} = (X, Y)$, where $\bold{A}_1$ and $\bold{A}_2$ are the vector potentials of particle 1 and 2 respectively. Notice that the third term in Eq.\eqref{6} is completely imaginary. This comes from the fact that it is  first order in time derivative, hence analytically continuing to imaginary time ($t\rightarrow e^{-i\theta}\tau$, $\theta =\pi/2$) makes it completely imaginary. The results obtained here are not restricted only to two dimensions. It is completely general and can be extended to $n$-dimensional Euclidean space.

By choosing the symmetric gauge vector potential for the two particles $\bold{A}_i = \frac{1}{2} \bold{B} \times \bold{x}_i, \thinspace i=1,2$, where $\bold{B} = B_{\perp}\hat{z}$, the Euclidean Lagrangian can be written as

\bea
\begin{split}
\mathcal{L}_{E}&= \frac{m_1}{2}\lvert \dot{\bold{x}}_1\rvert^2 + \frac{m_2}{2}\lvert \dot{\bold{x}}_2\rvert^2+  i\frac{eB_{\perp}}{2}\left( \dot{\bold{x}}_1\times\bold{x}_1 -  \dot{\bold{x}}_2\times\bold{x}_2 \right) \\ &+ \frac{1}{2}m_1\omega^2_1\bold{x}_1^2 + \frac{1}{2}m_2\omega^2_2\bold{x}_2^2+ V(\lvert \bold{x}_1-\bold{x}_2\rvert).
\label{6a}
\end{split}
\eea 
One can show that the results are independent of the choice of gauge. The Lagrangian in Eq.\eqref{6a} describes motion of two
coupled particles of opposite charges in the plane, in the presence of
a constant transverse external magnetic field and
an external potential.  These two particles interact with each other by a nonlinear potential $V(\lvert \bold{x}_1-\bold{x}_2\rvert)$  which has a metastable minimum. The magnetic field breaks the time reversal symmetry of the Lagrangian and the weak harmonic oscillator potentials break the spatial translation symmetry of the Lagrangian. However, we can restore spatial translation invariance up to a total derivative in the limit $\omega_1 = \omega_2 =0$. This will be the case at the end of the calculation in this paper. Hence, the total linear momentum is conserved  and the dynamics of the system cannot, therefore, change the position of the center of mass \cite{em1}. Notice that the presence of the magnetic field makes the Lagrangian strictly two-dimensional.
\section{Effective Action}
 We proceed to the effect of an external transverse magnetic field  on the tunneling of the particles out of a metastable state by following the method of Caldeira and Leggett\cite{cl}. The partition function is given by
\bea
\begin{split}
Z &=\int d\bold{x}_1 d\bold{x}_2 K(\bold{x}_1,\bold{x}_2,\beta),
\label{7}
\end{split}
\eea
where
\bea
\begin{split}
K(\bold{x}_1,\bold{x}_2,\beta)& =\int\mathcal{D}\bold{x}_1 \int  \mathcal{D}\bold{x}_2 \exp\left(-S_E\right),
\label{8}
\end{split}
\eea
and
\bea
\begin{split}
S_E &= \int_0 ^\beta d\tau \mathcal{L}_{E},
\label{9}
\end{split}
\eea
 $\beta = 1/T$ is the inverse temperature. The tunneling rate is proportional to  $\exp(-S_E ^c)$,
where the Euclidean classical action $S_E ^c$ is determined from the bounce solution of the equation  $\delta S_E = 0$,
in which the periodic boundary condition $\bold{x}_1(0)=\bold{x}_1(\beta)$ and  $\bold{x}_2(0)=\bold{x}_2(\beta)$ are required. 

We will set $\hbar =1$ throughout the calculation in this paper. Let us simplify the problem by taking $m_1=m_2=m$ and $\omega_1=\omega_2 =\omega^{\prime}$ and now introduce the following change of variables
\bea
\begin{split}
\bold{q} &= \bold{x}_1-\bold{x}_2,\quad \bold{r}&= \frac{\bold{x}_1+\bold{x}_2}{2}
\label{10}
\end{split}
\eea
where $\bold{q}$  is the position of particle 1 relative to particle 2 and $\bold{r}$ is the position vector of the center of mass of particles 1 and 2. The Lagrangian in the new coordinate system is of the form
\bea
\begin{split}
\mathcal{L}_{E}&= \frac{1}{2}\tilde{m}\left(\dot{ q_i}^2+ \omega^{\prime 2} {q_i}^2\right) + V(\lvert q_i\rvert)   \\ +&  \frac{M}{2}\left(\dot{r_i}^2 +\omega^{\prime 2}r^2_i\right) +   i {eB_{\perp}} \epsilon_{ij}\dot{r_i}q_j + \frac{ieB_{\perp}}{2}\frac{d}{d \tau} (\epsilon_{ij}q_i r_j) ,
\label{11}
\end{split}
\eea
where subscript $i=1,2$, $\tilde{m} = \frac{1}{2}m $ is the reduced mass and $M =2m$ is the total mass.  The last term in Eq.\eqref{11} is a total derivative and thus has no contribution to the classical equation of motion.  However, this term cannot in general be ignored when computing the quantum transition amplitude because it can generate phase terms in the Euclidean action that may, in principle, produce oscillations of the tunnelling amplitude on the applied field, but since we impose periodic boundary condition on the coordinates $q_i(\beta)=q_i(0)$ and $r_i(\beta)=r_i(0)$, the total derivative term integrates out from the action. Thus, it can be ignored from Eq.\eqref{11}.

The density matrix becomes
\bea
\begin{split}
K(q_i,r_i,\beta)& =\int_{q_i(0)=q_{i0}}^{q_i(\beta)=q_{i0} }\mathcal{D}q_i \int_{r_i(0)=r_{i0}}^{r_i(\beta)=r_{i0} } \mathcal{D}r_i \exp\left(-S_E\right).
\label{12}
\end{split}
\eea
 Exploiting the periodic boundary conditions on $q_i$ and $r_{i}$, one can expand these coordinates in terms of Fourier series \cite{uw, ss}:
\bea
\begin{split}
 r_i(\tau) &=\frac{1}{\beta}\sum_{n=-\infty}^{n=\infty}r_{in}e^{i\omega_n\tau}, \quad \text{etc,}
\label{13}
\end{split}
\eea
where $r_{-in} = r_{in}^{*}$ and $\omega_n = -\omega_{-n} = 2\pi n/\beta$ is the bosonic matsubara frequency.
The classical equation of motion for $r_i$ is
\bea
\begin{split}
&M\ddot{\bar{r_i}} +ieB_{\perp}\epsilon_{ij}\dot{\bar{q_j}}-M\omega^{\prime 2}\bar{r_i}=0.
\label{14}
\end{split}
\eea
Fourier transforming \eqref{14} we obtain
\bea
\begin{split}
 \bar{r}_{in} =-\frac{eB_{\perp}\omega_n \epsilon_{ij}q_{jn}}{M(\omega_n^2 +\omega^{\prime 2}) }.
\label{15}
\end{split}
\eea
For any path, we write
\bea
\begin{split}
 r_{i}(\tau) = \bar{r}_{i}(\tau) + y_{i}(\tau),
\label{16}
\end{split}
\eea
so $y_{i}(0) =y_{i}(\beta)=0$. The Fourier transform of the center of mass coordinate $r_{i}(\tau)$ action becomes
%
\begin{align}
\begin{split}
 \mathcal{S}_E ^{r}&= \frac{1}{\beta}\sum_{n=-\infty}^{n=\infty}\frac{1}{2}M(\omega_n^2 +\omega^{\prime 2})\lvert y_{in}\rvert^2 \\& + \frac{1}{\beta}\sum_{n=-\infty}^{n=\infty}\frac{(eB_{\perp}\omega_n)^2}{2M(\omega_n^2 +\omega^{\prime 2})}\lvert q_{in}\rvert^2,.
\label{18}
\end{split}
\end{align}
Notice that the linear term in $y_{in}$ vanishes by means of the equation of motion. $\mathcal{D}r_i=\mathcal{D}r_{in} =\mathcal{D}y_{in}$, the Gaussian integral over $y_{in}$ in Eqn.\eqref{12} is easily done, we finally obtain the effective action:
\bea
\begin{split}
S_{E}^{eff}&= \int_{0}^{\beta}  d \tau\left\lbrace\frac{\tilde{m}}{2}\dot{ q_i}^2 + V(\lvert q_i\rvert)\right\rbrace \\&-\frac{1}{2\beta}\sum_{n=-\infty}^{n=\infty}\mathcal{A}_n \lvert q_{in}\rvert^2,
\label{19}
\end{split}
\eea
where
\bea
\begin{split}
 \mathcal{A}_n&= -\left( \tilde{m} \omega^{\prime 2} +\frac{ (eB_{\perp}\omega_n)^2 }{M(\omega_n^2 +\omega^{\prime 2})}\right) .
\label{20}
\end{split}
\eea

The equivalent form of the Caldeira Leggett effective action \cite{cl} one can obtain from \eqref{19} is 
\bea
\begin{split}
S_{E}^{eff}&= \int_{0}^{\beta} d \tau \left\lbrace\frac{\tilde{m}}{2}\dot{ q_i}^2 + V(\lvert q_i\rvert)\right\rbrace \\&+\frac{1}{4 }\int_{0}^{\beta}d\tau \int_{0}^{\beta}d \tau^{\prime}\mathcal{A}(\tilde{\tau})\left[q_i(\tau)-q_i(\tau^{\prime})\right]^2.
\label{21a}
\end{split}
\eea
where $\tilde{\tau} = \tau -\tau^{\prime}$ and
\bea
\begin{split}
\mathcal{A}(\tau)= -\frac{M\omega_c^2}{\beta}\sum_{n=-\infty}^{n=\infty}\frac{\omega_n^2 e^{i\omega_n\tau}}{(\omega_n^2 +\omega^{\prime 2})},\\
\label{23}
\end{split}
\eea
$\omega_c = eB_{\perp}/M$ is the cyclotron frequency.
The first term in \eqref{20} is independent of $\omega_n$ and thus give an unnecessary delta function contribution to \eqref{23} which does not have any contribution to the effective action and hence can be neglected.  Further simplification of \eqref{23} yields
\bea
\begin{split}
\mathcal{A}(\tau)
\approx  \frac{M\omega_c^2}{\beta}\sum_{n=-\infty}^{n=\infty}\frac{\omega^{\prime 2}} {\left(\omega_n^2 +\omega^{\prime 2}\right)}e^{i\omega_n\tau}.
\label{23a}
\end{split}
\eea
 The above expression can now be summed easily by means of residue theorem or the summation formula \cite{ii,uw}, it is given by
 \bea
\begin{split}
\mathcal{A}(\tau) =\frac{M\omega_c^2}{2} \frac{\omega^{\prime}\cosh\left[\omega^{\prime}(\beta/2-\lvert \tau\rvert)\right]}{\sinh(\beta\omega^{\prime}/2)}.
\label{23b}
\end{split}
\eea
The effective action then becomes
\bea
\begin{split}
S_{E}^{eff}&= \int_{0}^{\beta} d \tau \left\lbrace\frac{\tilde{m}}{2}\dot{ q_i}^2 + V(\lvert q_i\rvert)\right\rbrace \\&+\frac{M\omega_c^2} {4} \int_{0}^{\beta}d\tau^{\prime} \int_{0}^{\beta}d \tau \frac{\omega^{\prime}}{2} \frac{\cosh\left[\omega^{\prime}(\beta/2-\lvert \tau\rvert)\right]}{\sinh(\beta\omega^{\prime}/2)}\\&\times\left[q_i(\tau)-q_i(\tau^{\prime})\right]^2.
\label{q}
\end{split}
\eea
%
%
In the limit $\omega^{\prime} \rightarrow 0$, Eqn.\eqref{23b} simplifies to 
\bea
\mathcal{A}{(\tau)}=&  {M \omega_c^2}/ {\beta},
\eea

and the action is thus
\bea
\begin{split}
S_{E}^{eff}&= \int_{0}^{\beta} d \tau \left\lbrace\frac{\tilde{m}}{2}\dot{ q_i}^2 + V(\lvert q_i\rvert)\right\rbrace \\&+\frac{M\omega_c^2} {4\beta} \int_{0}^{\beta}d\tau^{\prime} \int_{0}^{\beta}d \tau \left[q_i(\tau)-q_i(\tau^{\prime})\right]^2.
\label{21aa}
\end{split}
\eea
The effective actions in Eqn's.\eqref{q} and \eqref{21aa} are strictly two dimensional $(i=1,2)$ version of Eqn.(20) and Eqn.(13) in Ref.\cite{ad} for the case of finite pinning, no dissipation and no pinning, no dissipation respectively. However, in the present case the effective action depends on the relative coordinate of the two particles.
\section{LEGGETT'S PRESCRIPTION AT ZERO TEMPERATURE}

Let us now derive the zero temperature ($\beta\rightarrow \infty$) version of Eqn. \eqref{21aa} by  applying the Leggett's prescription \cite{al}. This prescription simply tells us that if the Fourier transform of the real time classical equation of motion is of the form
\bea
K(\omega)q_i(\omega) = -\left(\frac{dV}{dq_i}\right)(\omega),
\label{l2}
\eea
then the formula for the tunneling rate can be obtained from the effective action 
\bea
S_E^{eff} =\frac{1}{2\pi}\int_{-\infty}^{\infty} \frac{1}{2}K(-i\lvert\omega\rvert)\lvert\tilde{q_i}(\omega)\rvert^2 d \omega + S_{v}(\tilde{q_i}(\omega)),
\label{l2}
\eea
where
\bea
S_{v}(\tilde{q_i}(\omega))\equiv\int_{-\infty}^{\infty}d\tau V(q_i(\tau)),
\eea
and $\tilde{q_i}(\omega)$ is the Fourier transform of the imaginary-time trajectory. 
Now the real time classical equations of motion from \eqref{11} are
\bea
M(\ddot{r_i} + \omega^{\prime 2}r_i)-eB_{\perp}\epsilon_{ij}\dot{q}_j =0.
\label{l3}
\eea
\bea
\begin{split}
&\tilde{m}(\ddot{q_i} +\omega^{\prime 2}q_i)-eB_{\perp}\epsilon_{ij}\dot{r}_j=-\frac{dV}{dq_i}.
\end{split}
\label{l4}
\eea
Next, we perform the real time Fourier transform Eqn's.\eqref{l3} and \eqref{l4} at zero temperature (see Appendix) and solve for $q_{i}(\omega)$. The result is of the form
\bea
K(\omega)q_{i}(\omega) = -\left(\frac{dV}{dq_i}\right)({\omega}),
\label{l5}
\eea
where
\bea
\begin{split}
K(\omega)&= -\tilde{m}(\omega^2 -\omega^{\prime 2})+\frac{(eB_{\perp}\omega)^2}{M(\omega^2 -\omega^{\prime 2})}.
\label{l6}
\end{split}
\eea
Plugging \eqref{l6} into \eqref{l2}, we obtain
\bea
\begin{split}
S_E^{eff} &=\frac{1}{2\pi}\int_{-\infty}^{\infty} d\omega\frac{1}{2}\left[\tilde{m}(\omega^2 +\omega^{\prime 2})\right. \\&\left. + \frac{(eB_{\perp}\omega)^2}{M(\omega^2 +\omega^{\prime 2})} \right]\lvert \tilde{ q}_{i}(\omega)\rvert^2 + S_{v}( \tilde{ q}_{i}(\omega)).
\label{l7}
\end{split}
\eea
One can simply derive this equation from  \eqref{19} as $\beta \rightarrow \infty$ by Fourier transforming the kinetic term in \eqref{19}, and replacing $\omega_n$ by $\omega$ and the summation over $n$ by integration over $\omega$ with a normalization factor of $1/2\pi$. Therefore, we see that the results are consistent.

\section{Dissipative environment} 

In this section we shall consider the coupling of the Lagrangian in \eqref{6} to a thermal harmonic oscillators in two dimensions. Therefore classical dynamics of the system will be dissipative. The Euclidean Lagrangian we will consider is of the form
\bea
\begin{split}
\mathcal{L}_{E}&= \frac{m_1}{2}\lvert \dot{\bold{x}}_1\rvert^2 + \frac{m_2}{2}\lvert \dot{\bold{x}}_2\rvert^2+  i\frac{eB_{\perp}}{2}  \left( \dot{\bold{x}}_1\times{\bold{x}}_1 -  \dot{\bold{x}}_2\times{\bold{x}}_2 \right)\\
& + \frac{1}{2}m_1\omega^2_1\bold{x}_1^2 + \frac{1}{2}m_2\omega^2_2\bold{x}_2^2+ V(\lvert \bold{x}_1-\bold{x}_2\rvert)\\&+ \sum_{\alpha=1}^{N}\frac{m_{\alpha}}{2}\left[\dot{\bold{x}}\thinspace_{\alpha}^2 + \omega_{\alpha}^2\left(\bold{x}_{\alpha}-\bold{x}\right)^2\right],
\label{21d}
\end{split}
\eea 
where $\bold{x} = (\bold{x}_1,\bold{x}_2)$.

 In the absence of a transverse magnetic field and weak harmonic oscillator potentials, the Lagrangian corresponds to a two dimensional version of the one considered in Ref. \cite{em1} except for the coupling in $\bold{x}$ instead of $\bold{x}_2$. The Lagrangian is also similar to the one studied in Ref. \cite{pa1} where it was shown that the action can be mapped to a one-dimensional problem. However, this is not the case in Eqn.\eqref{21d}. We have assumed that both particles are coupled to a large environmental harmonic oscillators. Notice that the Lagrangian is still translational invariant up to a total derivative when $\omega_1=\omega_2 = 0$. Therefore translational invariance of the system will be restored by taking the limit $\omega_1=\omega_2 = 0$ at the end of the calculation.
 
 In order to obtain the effective action, we expand $\bold{x}_{\alpha}$ and $\bold{x}$ in a Fourier series:
 \bea
\begin{split}
 \bold{x}_{\alpha}(\tau) &=\frac{1}{\beta}\sum_{n=-\infty}^{n=\infty}\bold{x}_{\alpha n}e^{i\omega_n\tau},\\
 \bold{x}(\tau) &=\frac{1}{\beta}\sum_{n=-\infty}^{n=\infty}\bold{x}_{n}e^{i\omega_n\tau}, \quad\text{etc.}
\label{21d1}
\end{split}
\eea

Performing the Gaussian integration over $\bold{x}_{\alpha n}$ we obtain
 \bea
\begin{split}
K(\bold{x}_1,\bold{x}_2,\beta)& =\int \mathcal{D}\bold{x}_1(\tau) \int   \mathcal{D}\bold{x}_2(\tau) \exp\left(- S_E\right),
\label{21e}
\end{split}
\eea
where
\bea
\begin{split}
S_E &=\int_{0}^{\beta} d \tau\ \left[\frac{m}{2}(\dot{\bold{x}}\thinspace_1^2   +  \dot{\bold{x}}\thinspace_2^2 )+ i\frac{eB_{\perp}}{2}  \left( \dot{\bold{x}}_1\times{\bold{x}}_1 -  \dot{\bold{x}}_2\times{\bold{x}}_2 \right)\right. \\& \left.+ \frac{1}{2}(m\omega^{\prime 2}+\sum_\alpha m_{\alpha}\omega_{\alpha}^2)\bold{x}^2 + V(\lvert \bold{x}_1-\bold{x}_2\rvert)  \right] \\&-\frac{1}{\beta}\sum_{\alpha n}\frac{m_{\alpha}\omega_{\alpha}^4}{2(\omega_{n}^2+ \omega_{\alpha}^2)}\lvert \bold{x}_{n}\rvert ^2.
\label{21f}
\end{split}
\eea
We have set $m_1=m_2=m$ and $\omega_1=\omega_2 =\omega^{\prime}$ to arrive at this result. Next, we rewrite \eqref{21f} in terms of $q_{i}$ and $r_{i}$ using \eqref{10}, Fourier transform using \eqref{13} and use the fact that the Fourier coefficient $\lvert \bold{x}_{n}\rvert ^2 =\lvert \bold{x}_{1n}\rvert ^2 + \lvert \bold{x}_{2n}\rvert ^2$ where
\begin{align}
\bold{x}_{1n} = \bold{r}_{n}+\frac{1}{2}\bold{q}_{n},\quad \text{and} \quad \bold{x}_{2n} = \bold{r}_{n}-\frac{1}{2}\bold{q}_{n}.
\end{align}
Then the Gaussian integration over $\bold{r}_n$ can be easily done, and we obtain
\bea
\begin{split}
K(q_{i};\beta)& =\int \mathcal{D}q_{i}(\tau) \ \exp\left(- S_E^{eff}\right),
\label{21g}
\end{split}
\eea
where
\bea
\begin{split}
S_{E}^{eff}&= \int_{0}^{\beta}  d \tau\left(\frac{\tilde{m}}{2}\dot{q_{i}}\thinspace ^2  + V(\lvert q_{i}\rvert)\right) \\&+\frac{1}{4}\int_{0}^{\beta} d\tau\int_{0}^{\beta}d\tau^{\prime}\left(\mathcal{A}(\tilde{\tau})+\mathcal{B}(\tilde{\tau}) \right) \left[q_{i}(\tau)-q_{i}(\tau^{\prime})\right]^2
\label{21h}
\end{split}
\eea
The coefficients are
\begin{align}
\mathcal{A}(\tau)&= \frac{(eB_{\perp})^2}{M\beta}\sum_{n}    \frac{M\omega^{\prime 2} + 2\lambda_{n} \omega_{n}^{2}}{M (\omega_{n}^{2}+\omega^{\prime 2})  + 2\lambda_{n} \omega_{n}^{2}}e^{i\omega_n\tau } \label{solo} \\
\mathcal{B}(\tau) &= -\frac{1}{2\beta}\sum_{n} \omega_{n}^2\lambda_{n}e^{i\omega_n\tau }\label{solo1}\\
\lambda_n(\beta) &= \sum_{\alpha}\frac{m_{\alpha}\omega_{\alpha}^2 }{\omega_{\alpha}^2 +\omega_{n}^2}. 
\label{21j}
\end{align}
%

Constant terms independent of $\omega_n$ have been dropped since they give no contribution to \eqref{21h}. In general, Eqn.\eqref{solo} is difficult to sum unless one considers some limiting cases.

\section{Ohmic dissipation} 
 In line with Ref.\cite{em1}, we will assume that the effect of the oscillators on the motion of the particles result in the force of friction $\eta\dot{\bold{x}}$. This requires that the spectral density should be defined as
\bea
\begin{split}
 J(\tilde{\omega})= \frac{\pi}{2}\sum_{\alpha}m_{\alpha}\omega_{\alpha}^{3}\delta(\tilde{\omega}-\omega_{\alpha}).
\label{21k}
\end{split}
\eea
All the information concerning the effect of the environment on the dynamics of the particles is contained in $J(\tilde{\omega})$. The spectral function is frequently assumed to be of the form \cite{cl, ad}
\bea
\begin{split}
 J(\tilde{\omega})=  \eta\tilde{\omega}^s \exp\left(-\tilde{\omega}
 /\omega_c\right),
 \label{21l}
\end{split}
\eea
up to a frequency cutoff $\omega_c$, where $s>1$ is the super-Ohmic case, $s=1$ is the Ohmic case, and $0\leq s < 1$ is the sub-Ohmic case. In this section, we will consider only the case of Ohmic dissipation  with $\omega_c \rightarrow \infty$. 
Using the definition of the spectral function \eqref{21k} we have
\bea
\begin{split}
 \lambda_{n} = \frac{2}{\pi}\int_{0}^{\infty}\frac{d\tilde{\omega}}{\tilde{\omega}} \frac{J(\tilde{\omega})}{\tilde{\omega}^2 +\omega_{n}^2}= \frac{\eta}{\lvert \omega_n\rvert}.
 \label{21m}
\end{split}
\eea
The second equality follows from \eqref{21l} for the Ohmic case. The effective action \eqref{21h} in this case becomes 

\bea
\begin{split}
S_{E}^{eff}&= \int_{0}^{\beta}  d \tau\left(\frac{\tilde{m}}{2}\dot{q_{i}}\thinspace ^2  + V(\lvert q_{i}\rvert)\right) \\&+\frac{1}{4}\int_{0}^{\beta} d\tau \int_{0}^{\beta}d\tau^{\prime} \left[\mathcal{A}(\tilde{\tau}) + \mathcal{B}(\tilde{\tau})\right]    \left[q_{i}(\tau)-q_{i}(\tau^{\prime})\right]^2 
\label{s9}
\end{split}
\eea
where
\begin{align}
\mathcal{A}(\tau)&= \frac{(eB_{\perp})^2}{M\beta}\sum_{n}    \frac{M\omega^{\prime 2} + 2\eta \lvert\omega_n\rvert}{M (\omega_{n}^{2}+\omega^{\prime 2})  + 2\eta \lvert\omega_n\rvert}e^{i\omega_n\tau },
\label{s10}\\
\mathcal{B}(\tau)&= -\frac{1}{\beta}\sum_n \frac{\eta \lvert\omega_n\rvert}{2}e^{i\omega_n\tau }
\label{s11}.
\end{align}

Let us consider the limit of very strong dissipation $\eta \gg M$, setting $\omega^{\prime} =0 $ we have
\begin{align}
\frac{2\eta \lvert\omega_n\rvert}{M \omega_{n}^{2}   + 2\eta \lvert\omega_n\rvert} \approx 1-\frac{M\lvert\omega_n\rvert}{2\eta}
\label{s14}.
\end{align}
Hence the effective action becomes
\bea
\begin{split}
S_{E}^{eff}&= \int_{0}^{\beta}  d \tau\left(\frac{1}{2}\tilde{m}\dot{q_{i}}\thinspace ^2  + V(\lvert q_{i}\rvert)\right) \\&+\frac{\eta_{eff}}{8\pi}\int_{0}^{\beta} d\tau^{\prime}\int_{0}^{\beta}d\tau \frac{\left[q_{i}(\tau)-q_{i}(\tau^{\prime})\right]^2}{(\beta\slash\pi)^2\sin^2(\pi\tilde{\tau}\slash \beta)},
\label{21h8}
\end{split}
\eea
where
\bea
\eta_{eff} = \frac{(eB_{\perp})^2}{\eta} +\eta.
\eea
In the limit $B_{\perp}=0$, the action corresponds to a two dimensional version of Eq.(44) in Ref.\cite{em1} for $m_1=m_2$. 

At $T=0$, we apply the Leggett's prescription outline above. The real time classical equations of motion for the case in which the motion of the two identical particles are damped by Ohmic friction with constant $\eta$ are
\begin{align}
m\ddot{x}_i^1 -eB_{\perp}\epsilon_{ij}\dot{x}_j^1 +\eta\dot{x}_i^1 + m\omega^{\prime 2}{x}_i^1=-\frac{\partial V}{\partial x_i^1},
\label{s}\\
m\ddot{x}_i^2 +eB_{\perp}\epsilon_{ij}\dot{x}_j^2 +\eta\dot{x}_i^2 +m\omega^{\prime 2}{x}_i^2=-\frac{\partial V}{\partial x_i^2}
\label{s1}.
\end{align}
Substituting $x_i^1$ and $x_i^2$ in terms of $r_i$ and $q_i$ using the inverse transformation of \eqref{10}, we obtain, after adding and subtracting the resulting equations 
\bea
\begin{split}
2m\ddot{r}_i  -eB_{\perp}\epsilon_{ij} \dot{q}_j +2\eta\dot{r}_i +2m\omega^{\prime 2}{r}_i &= 0,\\
\frac{1}{2}m\ddot{q}_i -eB_{\perp}\epsilon_{ij}\dot{r}_j   + \frac{1}{2}\eta\dot{q}_i + \frac{m\omega^{\prime 2}{q}_i}{2}&=-\frac{\partial V}{\partial q_i}.
\label{s1}
\end{split}
\eea
Fourier transforming \eqref{s1} and solving for $q_{i}(\omega)$ we obtain
\bea
K(\omega)q_{i}(\omega) =-\frac{\partial V}{\partial q_i}(\omega),
\eea
where $K(\omega)$ in this case is given by
\begin{align}
K(\omega)&= -\tilde{m}(\omega^2 -\omega^{\prime 2}) +\frac{(eB_{\perp}\omega)^2}{M(\omega^2-\omega^{\prime 2})- 2i\eta \omega}\\& + \frac{i\eta \omega}{2}.\nonumber
\end{align}
The effective action at $T =0$ is thus
\bea
\begin{split}
S_{E}^{eff} &=\frac{1}{2\pi}\int_{-\infty}^{\infty} d\omega\frac{1}{2}\left[\tilde{m}(\omega^2 + \omega^{\prime 2}) + \frac{\eta\lvert\omega\rvert}{2 }\right.\\&\left. +\frac{(eB_{\perp}\omega)^2}{M(\omega^2+ \omega^{\prime 2}) +2\eta \lvert\omega\rvert}\right]\lvert \tilde{ q}_{i}(\omega)\rvert^2 + S_{v}( \tilde{ q}_{i}(\omega)).
\label{s3}
\end{split}
\eea
 
\begin{figure}[h!]
\centering
\includegraphics[scale=0.25]{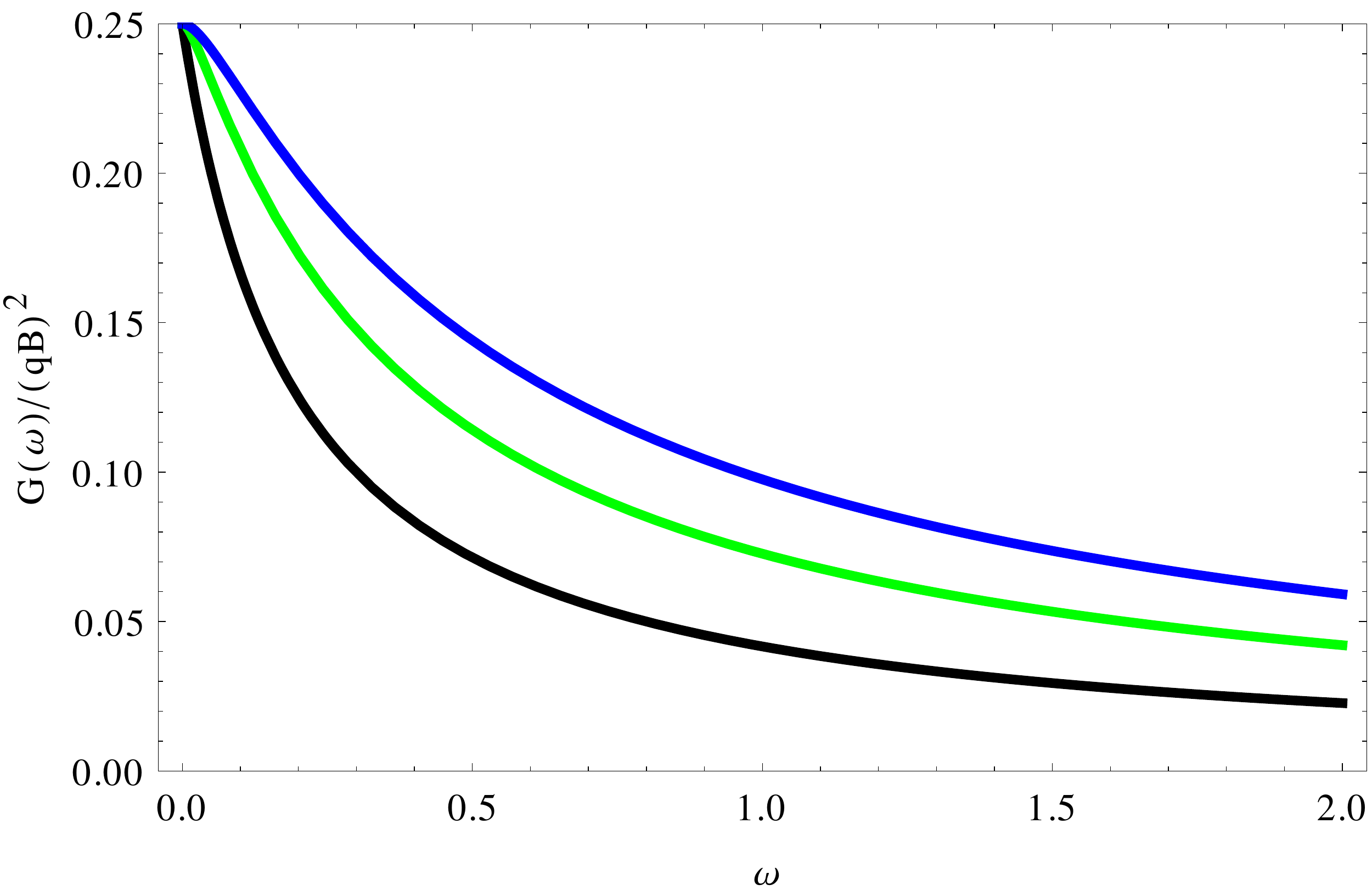}
\caption{(Color online): Plot of $ \mathcal{G}(\omega)$ as a function of $\omega$ for $M =1$. $\eta = 0.1$, $\omega^{\prime} =0$ $(\text{black})$,  $\eta = 0.2$, $\omega^{\prime} =0.1$ $(\text{green})$,  $\eta = 0.3$, $\omega^{\prime} =0.2$ $(\text{blue})$.}  \label{fig4}
\end{figure}

Fourier transforming back to imaginary time domain (see Appendix) we obtain
\bea
\begin{split}
S_{E}^{eff}= &\int_{-\infty}^{\infty}  d \tau\left(\frac{1}{2}\tilde{m}\dot{q_{i}}\thinspace ^2  + V(\lvert q_{i}\rvert)\right)   \\&+ \frac{\eta}{8\pi}\int_{-\infty}^{\infty} d\tau^{\prime}\int_{-\infty}^{\infty}d\tau \thinspace \frac{\left[q_{i}(\tau)-q_{i}(\tau^{\prime})\right]^2 }{\lvert \tau-\tau^{\prime}\rvert^2} \\&+
\int_{-\infty}^{\infty} d\tau^{\prime}\int_{-\infty}^{\infty}d\tau \thinspace \mathcal{G}(\tau -\tau^{\prime})\left[q_{i}(\tau)-q_{i}(\tau^{\prime})\right]^2,
\label{s4}
\end{split}
\eea
where
\begin{align}
\mathcal{G}(\tau -\tau^{\prime})&=   \int_{-\infty}^{\infty} \frac{ d \omega}{2\pi}\mathcal{G}(\omega)e^{i\omega(\tau -\tau^{\prime})}, 
\label{s10}
\end{align}
and the Fourier coefficient is
 \bea
 \mathcal{G}(\omega)= \frac{(eB_{\perp})^2}{4M }\frac{M \omega^{\prime 2}+2\eta \lvert\omega\rvert} {M (\omega^{2}+\omega^{\prime 2})  + 2\eta \lvert\omega\rvert}.
 \label{s11}
 \eea
 As usual, we have dropped constant terms independent of $\omega$. The plot of $ \mathcal{G}(\omega)$ as a function of $\omega$ is shown in Fig.\eqref{fig4}. For $\omega^{\prime} =0$, the integration in Eq.\eqref{s10} gives in the large $\tau-\tau^{\prime}$ limit \cite{ad, pa1}
 \bea
 \mathcal{G}(\tau -\tau^{\prime})= \frac{(eB_{\perp})^2}{8\pi\eta}\frac{1}{\lvert \tau -\tau^{\prime}\rvert^2}.
 \eea
Plugging this expression into Eq.\eqref{s4} we recover a two-dimensional version of Eq.(18) in Ref. \cite{ad} for the case of no pinning and finite dissipation. 

\section{Conclusions} We have studied  the macroscopic quantum tunneling of two coupled particles of identical masses but opposite charges in the plane, in the presence of a constant transverse external magnetic field and an external potential whose interaction potential allows for a metastable state. We showed that the effect of the magnetic is to suppress the tunneling of the particles out of a metastable state and also the effective action remains two dimensional unlike a one dimension version obtained in Ref. \cite{pa}. 

We further coupled the system to a thermal bath of harmonic oscillator and showed that in the limit of strong dissipation, there is an effect of the magnetic field to the effective action. In the limit of zero magnetic field, we reproduced a two dimensional version of the results obtained in Ref. \cite{em1} for $m_1=m_2$, which also coincides with the results of Caldeira and Leggett \cite{cl} in two dimensions in terms of the reduced mass. The results obtained in the paper are as a consequence of the conservation of total linear momentum as was shown in Ref.\cite{em1}.
These results can be applied to the metastable states of the molecules of fluid and solids in a transverse magnetic field and also to coupled superconductor vortex tunneling in two dimensions.
 
\section{Acknowledgments} I would like to thank Manu Paranjape for his support.  
\section{Appendix}
The real time Fourier transform is defined as
\bea
\begin{split}
 q_{i}(t) &= \frac{1}{2\pi}\int_{-\infty}^{\infty}d\omega \thinspace q_{i}(\omega)e^{i\omega t},\quad \text{etc}
\label{13a}
\end{split}
\eea
and the imaginary time Fourier transform is defined as
\bea
\tilde{q_{i}}(\omega) = \int_{-\infty}^{\infty}q_{i}(\tau)e^{-i\omega\tau}d\tau
\eea
Fourier transforming Eq .\eqref{s3} we have for the potential term
\bea
S_{v}( \tilde{ q_{i}}(\omega)) = \int_{-\infty}^{\infty}d\tau V(q_{i}(\tau)).
\label{s5}
\eea
The first term in Eq.\eqref{s3} gives
\begin{align}
S_1 =&\int_{-\infty}^{\infty}d\tau\int_{-\infty}^{\infty}
d\tau^{\prime}\frac{1}{2}\tilde{m}\dot{q_{i}}(\tau)\dot{q_{i}}(\tau^{\prime})\frac{1}{2\pi}\int_{-\infty}^{\infty} d\omega e^{i\omega(\tau-\tau^{\prime})}\\&
= \int_{-\infty}^{\infty}\frac{1}{2}\tilde{m}\dot{q_{i}}(\tau)^2 \nonumber
\end{align}
Note the contribution from $\omega^{\prime}$ gives a delta function and hence a factor of $q_{i}(\tau)^2$ which can then be absorbed in the potential Eq.\eqref{s5}.

The second term in Eq.\eqref{s3} gives
\begin{align}
S_2 =&\frac{1}{8\pi} \int_{-\infty}^{\infty}d\tau\int_{-\infty}^{\infty}
d\tau^{\prime}\int_{-\infty}^{\infty} d\omega \thinspace \eta\lvert\omega\rvert e^{i\omega(\tau-\tau^{\prime})}q_{i}(\tau)q_{i}(\tau^{\prime})
\label{s6}  
\end{align}
Now we use the fact that
\bea
\eta\lvert\omega\rvert =\frac{2\eta}{\pi}\int_{0}^{\infty} du \frac{\omega^2}{u^2 + \omega^2}.
\label{s7}
\eea
Plugging \eqref{s7} into \eqref{s6} and performing the contour integration over $\omega$ and subsequently integration over $u$ we obtain
\begin{align}
S_2 =&-\frac{1}{4\pi}\int_{-\infty}^{\infty}
d\tau^{\prime}\int_{-\infty}^{\infty}d\tau\frac{q_{i}(\tau)q_{i}(\tau^{\prime})}{\lvert \tau-\tau^{\prime}\rvert^2}
\label{s8}  
\end{align}

The last term in Eq.\eqref{s3} is simply
\bea
S_3= 2\int_{-\infty}^{\infty}
d\tau^{\prime}\int_{-\infty}^{\infty}d\tau \mathcal{G}(\tau-\tau^{\prime}) q_{i}(\tau)q_{i}(\tau^{\prime})
\eea
We can use the fact that 

\[q_{i}(\tau)q_{i}(\tau^{\prime}) = \frac{1}{2}\left[q_{i}(\tau)^2 + q_{i}(\tau^{\prime})^2 - (q_{i}(\tau ) -q_{i}(\tau^{\prime}))^2\right]\]
to arrive at Eq.\eqref{s4}.

\end{document}